# Topographic Deep Artificial Neural Networks (TDANNs) predict face selectivity topography in primate inferior temporal (IT) cortex

**Hyodong Lee (hyo@mit.edu)**
Department of Electrical Engineering and Computer Science, Massachusetts Institute of Technology
Cambridge, Massachusetts 02139

**James J. DiCarlo (dicarlo@mit.edu)**
McGovern Institute for Brain Research and Department of Brain and Cognitive Sciences,
Massachusetts Institute of Technology
Cambridge, Massachusetts 02139 United States

**Abstract:**

**Deep convolutional neural networks are biologically driven models that resemble the hierarchical structure of primate visual cortex and are the current best predictors of the neural responses measured along the ventral stream. However, the networks lack topographic properties that are present in the visual cortex, such as orientation maps in primary visual cortex and category-selective maps in inferior temporal (IT) cortex. In this work, the minimum wiring cost constraint was approximated as an additional learning rule in order to generate topographic maps of the networks. We found that our topographic deep artificial neural networks (ANNs) can reproduce the category selectivity maps of the primate IT cortex.**

**Keywords: Deep neural network; Topography; Visual cortex;**

## Introduction

Abundant evidence suggests that the primate visual cortex consists of hierarchically structured cortical areas. However, the mechanisms of the topographic development of the areas are not well understood. One hypothesis is that the wiring cost derives the spatial organization of the brain network. Koulakov and Chklovskii (2001) have shown that a simple mathematical model with minimum wiring cost can reproduce the orientation preference maps in primary visual cortex (V1).

Recent work has shown that particular deep artificial neural network (ANN) models are the best predictors of neurons at multiple levels of the ventral stream and its supported core object recognition behavior. However these models are not spatially mapped to cortical tissue, which means that they cannot yet be used to compare with spatial maps of the ventral visual stream. In this work, we tested if the wiring cost minimization approach, applied to deep ANNs, can potentially explain the topographic maps found in primate IT cortex. We here simply focused on assessing the natural emergence of face patches in such models.

## Methods

Instead of directly minimize the wiring cost, we approximated the wiring cost as the response profile similarity of each pair of neurons. That is, we reasoned that, the more similar the profile, the closer together those two neurons should be to keep wiring costs low. We believe that this notion is consistent with observations that the pair-wise neuronal correlations in visual cortex decrease as a function of the distance of the neurons on the cortical surface (Smith & Kohn, 2008; Smith & Sommer, 2013). The specific cost function we used was derived from the neural recordings in macaque IT cortex (Figure 2).

The Topographic Deep Artificial Neural Networks (TDANNs) in this work are adaptations of the Alexnet (Krizhevsky, Sutskever & Hinton, 2012) architecture. The main difference is that we aimed to create a tissue map (2D layout) of the artificial neurons in the first fully connected (fc6) layer. We chose Alexnet as we take that to be one of the baseline models of the primate IT neural population (Figure 1B). Note that it is not simply a matter of placing the artificial neurons into a tissue map – the tuning of all the model parameters must be done under the spatial costs constraint. To do this,, each network unit was initially assigned a random position in a two-dimensional surface (the "tissue map"). While being optimized for image classification task, the networks were penalized if the response profile of the units did not follow the derived response profile. The networks were trained on the ILSVRC-2012 dataset, which contains 1.2 million images from 1000 object classes (Deng, Dong, Socher, Li, Li & Fei-

Fei, 2009). To test reproducibility, 10 networks were trained with different parameter initializations.

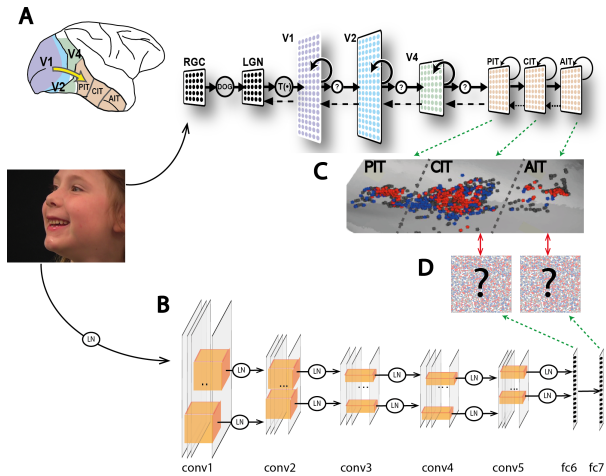

Figure 1: TDANNs as models of visual cortex. (A) The ventral visual pathway consists of hierarchically structured cortical areas (Yamins & Dicarlo 2016). (B) TDANNs are deep artificial neural networks simultaneously optimized for object recognition tasks and topographic constraints. (C,D) The topographic maps of TDANNs (D) may reproduce different functional maps in IT cortex (C), i.e. face selectivity maps.

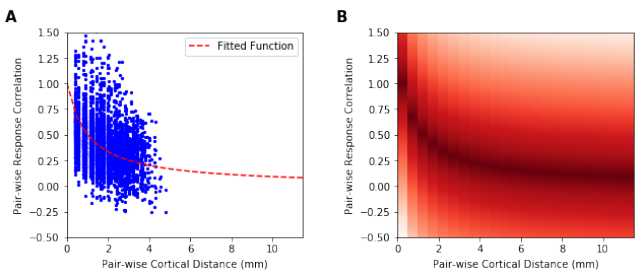

Figure 2: Response profile of IT and model neurons. (A) The blue dots represent the distribution of macaque IT neurons. The red line is the fitted function to the data. (B) The target response profile for the model neurons. The darker color means the less penalization.

## Results

We investigated the face-selectivity response maps of the model by presenting naturalistic visual stimulus sets (Figure 3A). The face selectivity of each neuron was measured as d' for face over non-face objects (Aparicio, Issa & DiCarlo, 2016). As shown in Figure 3B, the simulated tissue maps of the models reproduced the clustering of face-selective neurons, similar to the middle face patches (MFPs) in macaque IT cortex (Aparicio et al., 2016).

We also estimated the spatial profile of the "purity" of the face clusters. The purity of each 0.5mm x 0.5mm grid was computed as a percentage of face-selective neurons. Then, the grid with the highest purity value was identified as the center of the cluster. The fall-off characteristics of the purity curve were found to be similar to those of MFPs (Aparicio et al., 2016), as shown in Figure 3C.

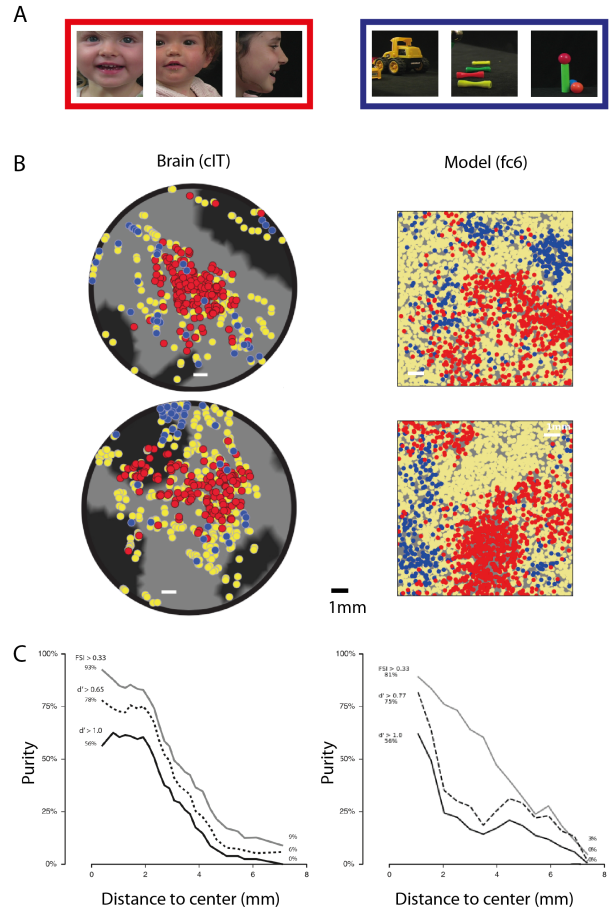

Figure 3: TDANNs can reproduce the face selectivity maps. (A) Sample face (red) and object (blue) stimuli. (B) Face selectivity map of brain (left) and model (right). Red or blue neurons responded preferentially to faces or objects. (C) Face cluster purity curve of brain (left) and model (right)

## Conclusion

In this work, we demonstrate that these new topographic deep ANNs (TDANNs) naturally and automatically produce face vs. object selectivity topography that is similar to that found in primate IT cortex. This suggests that the simple wiring cost minimization may derive the development of the topographic structure of the visual cortex. As a future direction, these models can be used to study the development of ventral stream spatial organization and its dependence on different types of visual experience.

## Acknowledgments

This research was supported by Samsung Scholarship, Intelligence Advanced Research Projects Agency (IARPA), and US National Eye Institute grants R21-EY025863. (J.J.D.).